\documentclass[twocolumn]{aastex62}

\usepackage{lineno}

\begin{document}

\title{IMPLICATIONS ON SPATIAL MODELS OF INTERSTELLAR GAMMA-RAY INVERSE-COMPTON EMISSION FROM SYNCHROTRON EMISSION STUDIES IN RADIO AND MICROWAVES}

\author[0000-0002-0786-7307]{E. Orlando}
\affil{Kavli Institute for Particle Astrophysics and
Cosmology, W.W. Hansen Experimental Physics Laboratory, Stanford University \\
CA 9305, Stanford, USA; \\
eorlando@stanford.edu}

\begin{abstract}

Cosmic rays interacting with gas and photon fields in the Galaxy produce interstellar gamma-ray emission (IGE), which accounts for almost 50\% of the photons detected at gamma-ray energies. Models of this IGE have to be very accurate for interpreting the high-quality observations by present gamma-ray telescopes, such as {\it Fermi} Large Area Telescope (LAT).
Standard models of IGE, used as reference models for analyses of the {\it Fermi} LAT data, show spatial discrepancies with respect to the data, underlining the necessity of more realistic models. 
The same CR electrons that produce the inverse-Compton component of the IGE produce also interstellar synchrotron emission observed in radio and microwave.
However, present standard models do not take advantage of results coming from studies of this interstellar synchrotron emission. \\
Accounting for such results, in this work we show how they affect the calculated spatial maps of the large-scale inverse-Compton component of the IGE, which are usually used in studies of Fermi LAT data. \\
It is found that these results significantly affect these spatial model maps even at a 60$\%$ level. In particular, propagation models based on synchrotron studies produce a more peaked inverse-Compton emission in the inner Galaxy region with respect to the standard models used to analyze Fermi LAT data.
The conclusion is that radio and microwave observations can be included in a multifrequency self-consistent approach for a more accurate modeling of the IGE finalized to a physical comprehensive interpretation of gamma-ray data and its present unexplained features. 
Model parameters are provided, which supply a more realistic basis for high-energy gamma-ray studies. \\

\end{abstract}


\section{Introduction} \label{sec:intro}
The interstellar gamma-ray emission (IGE) produced by Galactic Cosmic Rays (CRs) is a dominant component of the gamma-ray sky from a few tens of MeV to hundreds of GeV \citep[e.g.][]{Strong2004, diffuse1, diffuse2}. This emission is produced by interactions of CRs with photons and gas in the interstellar medium \citep{Strong2007}. 
It was detected already in the 70's with the Small Astronomy Satellite (SAS) and COS-B that observed the entire sky above 50 MeV. Since the era of the Compton Gamma-Ray Observatory (CGRO) the IGE has been carefully studied and modeled for one of two reasons: 1) IGE as confusing foreground for many analyses; 2) IGE as a useful tool for understanding CRs and the interstellar medium. Almost 50\% of the photons detected by {\it Fermi} Large Area Telescope \cite[LAT:][]{Atwood} is due to this IGE \citep{diffuse2}.
Nowadays, uncertainties in IGE models are limiting our understanding of the gamma-ray sky, especially for dark matter searches \citep[e.g.][and reference therein]{Linden,Calore,Carlson, Hooper, P8IG} and for source detection \citep[e.g.][]{GK, BK, Bartels, DiMauro}. Indeed, disentangling possible dark matter signals, or low-significance sources, from pure IGE is very challenging and needs realistic models. 
A detailed study of the IGE from the whole Galaxy was performed on a grid of 128 propagation models \citep{diffuse2} using {\it Fermi} LAT data. However, this study used standard reacceleration models only, which neglect synchrotron studies \citep[e.g.][]{Strong2011}. These models investigated different 2D CR source distributions, gas parameters, and propagation halo sizes. Even though all models provide a good agreement with data, two issues were identified. First, no best-fit model or a set of models could be identified for the whole sky. However, data prefer a large CR propagation halo size, and additional gas or CRs in the outer Galaxy. Second, many large-scale structures, such as the so-called Fermi Bubbles \citep{Su,FermiBubbles}, the inner Galaxy excess, and the outer Galaxy excess \citep{outer2010,outer2011} emerged in the observations as excesses over the adopted model. 
Moreover, many less structured excesses and even many dips with respect to present models are also shown all over the sky in the high-quality data by the {\it Fermi} LAT. Such excesses and dips are similar in intensity to the GeV excess seen in the Galactic center \citep{P8IG}. 
Usual analyses of the {\it Fermi} LAT data are based on IGE model maps from propagation models that are then fitted to the data. Forefront recent studies use alternative methods such as hydrodynamical simulations for the gas component \citep[e.g.][based on \cite{Pohl}]{Macias} and SkyFACT (Sky Factorization with Adaptive Constrained Templates) for modeling the gamma-ray emission with adaptive templates and penalized likelihoods \citep{Storm}. However they still rely on inverse-Compton maps from propagation models as input. \\
Hence, beside their quite good agreement, present models based on CR propagation and interaction codes such as GALPROP  \citep{Moskalenko98a,Strong2007,Vla,Orlando2013,Moskalenko2015}, are not able to describe in the entire sky the more and more precise observations of the diffuse emission, thereby challenging our understanding of the gamma-ray sky and its features. 
Improvements of these models, which are still the official reference models for studies of diffuse emission, will impact any research field that makes use of the IGE models, such as detection of sources, searches for dark matter, studies on the extragalactic emission, etc. This underlines the necessity of more accurate and realistic IGE models\footnote{A different approach for disentangling emission components in gamma rays independently from propagation models is followed for example by \cite{Selig}.}.\\
In the meantime, important updates on CR propagation models from studies of the interstellar synchrotron emission in radio and microwaves \citep{Strong2011, Orlando2013,PlanckBfield, O2018} have been obtained using the GALPROP code.
It is important to note that CR propagation models based on \cite{diffuse2}, and still used in most gamma-ray studies, do not include these results.
{\it The first result} from these studies is the derivation of the 3D Galactic magnetic field (B-field) distribution and intensity from synchrotron observations in radio and microwaves. 
Updated B-field models include both large-scale ordered and random components, with 3D spiral disc and halo constituents, and updated B-field intensities.
{\it The second result} is that standard reacceleration models as in \cite{diffuse2} are strongly disfavored by radio observations, preferring no, or very low, reacceleration instead. This is caused by the large amount of secondary electrons and positrons produced by reacceleration models at $\sim$ GeV energies. This overproduces synchrotron emission below a few tens of MHz that overestimates synchrotron data (models without reacceleration instead fit well synchrotron data). Despite this issue, reacceleration models have been the standard models for gamma-ray analyses so far.
Both results from synchrotron studies significantly affect the spatial distribution of primary and secondary electrons and positrons and their energy losses, and consequently also the spatial distribution of the calculated gamma-ray Inverse-Compton (IC) emission.  
Our recent work \citep[][hereafter O2018]{O2018} investigates the interstellar emission for the first time simultaneously in radio frequencies and in gamma-ray energies consistently, including constraints from latest CR direct measurements. While the effects on the calculated gamma rays regarding the spectrum were addressed in our previously mentioned work, here we report on the effects on the calculated gamma rays regarding the spatial distribution, i.e. on the maps used for gamma-ray analyses. 
This study adds on recent investigations on effects of 3D models of source distribution, gas and interstellar radiation field, which affect the modeling of the gamma-ray emission at several percent level \citep{Gulli}.\\
Here we show the spatial effects on IGE IC model maps when accounting for model constraints coming from the study of interstellar synchrotron emission observed in radio and microwave.  
These constraints significantly affect standard CR propagation models of IC emission and can help in providing insights on some physically unexplained features seen at gamma-ray energies.

\section{Method} 

\subsection{A multifrequency approach}\label{sec:concepts}
One of the major IGE components is the IC emission from CRs electrons and positrons on the optical and infrared interstellar radiation field, and on the cosmic microwave background. Direct observations of this emission is limited by contamination from the other emission  components (pion decay emission by CR protons and heavier nuclei on the gas, bremsstralhung emission by CR electrons and positrons on the gas), in addition to other diffuse emission components, such as the extragalactic background and the contribution of unresolved sources. 
Fortunately, the same CR electrons and positrons responsible for the IGE IC are also responsible for the interstellar emission seen at the opposite wavelengths, in radio and microwaves. In fact, this radio and microwave emission is produced by primary and secondary CR electrons and positrons spiraling in the Galactic B-field. Observations of this synchrotron emission encode information on B-fields, on CR electrons and positrons, and, hence, on propagation models. 
Our approach, pioneered in O2018, is to use studies of synchrotron emission in a consistent way to obtain information on propagation models used for gamma-ray studies.

\subsection{Significant results of recent synchrotron studies}
We summarize here the state of the art on studies of the interstellar synchrotron emission limited to those results that can affect the modeling of the IC component of the IGE.

Interstellar synchrotron emission is the most prominent diffuse component at low frequencies (below few GHz) measured by ground based radio instruments, and it is an important component in microwaves.
Observations of the interstellar synchrotron emission in the radio band from a few MHz to tens of GHz
were used to constrain CRs and propagation models by \cite{Strong2011}, finding that models with no reacceleration fit best synchrotron spectral data. This approach was followed by other similar
works \citep{Jaffe2011, dragon}. More recently, \cite{Orlando2013} investigated
the spectral and spatial distribution of the synchrotron emission in temperature
and polarization for the first time in the context of CR propagation
models. Various CR source distributions, CR propagation
halo sizes, propagation models (e.g. pure diffusion and diffusive-reacceleration
models), and 3D B-field formulations were studied against
synchrotron observations, highlighting degeneracies among the parameters used in the modeling. 
Additionally, in radio, as in gamma-ray energies, models with flat CR source distribution and large halo size were preferred. The best propagation model was identified and the best 3D B-field models and their intensities for ordered and random components were obtained. These models have been used for producing the low frequency foreground component maps released by the Planck Collaboration \citep{CompPlanck}.

Because the CRs responsible for the radio emission are also responsible for producing the gamma-ray emission, very recently we took advantage of this property with the aim of constraining CRs and propagation models by
looking at the interstellar emission in radio and gamma-ray energies
simultaneously (O2018). 
This approach provides a handle on both sides
of the electromagnetic spectrum in understanding CRs, thereby
leaving less room to uncertainties. In that work we also updated the B-field intensities for ordered and random components based on synchrotron Planck data \citep{LFPlanck,CompPlanck}, recent reprocessed 408 MHz map \citep{haslam1981, Remazeilles}, and by using the latest CR measurements with AMS02 \citep{AMS02, AMSele}. These B-field parameters are also updated with respect to the extensive work in \cite{PlanckBfield}, which, even though based on GALPROP models, instead used an early CR electron spectrum based on less precise {\it Fermi} LAT measurements \citep{fermi_ele}.
By studying the local interstellar spectrum we found (O2018) that both gamma-ray data and radio/microwave data prefer the spectrum of pure diffusion models with respect to standard reacelleration models, confirming earlier results by \cite{Strong2011} using radio/microwave data only. This is due to the high positron density $\sim$~1~GeV produced in the usual reacelleration models, that is in tension with the synchrotron spectrum at the low frequencies and also with the gamma-ray spectrum below $\sim$~1~GeV. 
Hence, both synchrotron (radio and microwave) spectral data and gamma-ray spectral data challenge standard re-acceleration models, in favor of diffusion models without or with small reacceleration. \\
The goal of this study is to present the outcome in the IC spatial model maps of using: 1) updated B-field models that fit synchrotron data; 2) propagation models with no reacceleration as supported by synchrotron and gamma-ray spectral data.

\section{Modeling}

This section describes how we build the CR propagation models and associated IC gamma-ray models used in this work.

\subsection{The GALPROP code}\label{sec:galprop}
CR propagation models and associated interstellar emission are computed by using the GALPROP code (see references above). GALPROP  is a numerical code for modeling the propagation of CRs in the Galaxy and for calculating the associated diffuse emissions. Other CR propagation codes are, for example, DRAGON \citep{Evoli}, PICARD \citep{Picard}, USINE \citep{Usine}. GALPROP solves the transport equation for all required CR species, for given CR source distribution, B-field, gas, and interstellar radiation field models. The propagation equation is solved numerically on a user-defined spatial grid in 2D or in 3D, and energy grid. It takes into account diffusion, convection, energy losses, ionization, and diffusive reacceleration processes. Secondary CRs produced by collisions in the gas, and decay of radiative isotopes are included. GALPROP is officially used by the {\it Fermi} LAT Collaboration \cite[e.g.][and reference therein]{diffuse1, diffuse2,  IEM, P8IG}. Recently, extensions of GALPROP for modeling the Galactic radio and microwave emission have been presented \citep{Orlando2013}. Starting with \cite{Strong2011} it has been extended to include calculations of radio temperature, polarization, absorption, and free-free emission, to model the interstellar emission in a consistent way from radio to gamma rays.

\subsection{CR propagation models}

Four CR propagation models are used. All models fit latest CR leptonic and hadronic measurements by Voyager I \citep{Voyager} and by AMS02 \citep{AMS02, AMSele}.

\startlongtable
\begin{deluxetable}{ccc}
\tablecaption{Model parameters \label{table1}}
\tablehead{
\colhead{\bf{Model}} & \colhead{\bf{B-field}} & \colhead{\bf{Propagation}} \\
\colhead{\bf{name}} & \colhead{\bf{(ord, ran)$^{a}$}} & \colhead{\bf{model}} \\
}
\startdata
PDDE$^{b}$ & updated & no reacceleration\\
& (2.7, 4.9) & \\
PDDEBold$^{c}$ & old & no reacceleration\\
& (-, 5) & \\
DRE$^{c}$ & updated & with reacceleration \\
& (2.7, 4.9) &  \\
DRE$\_$comb$^{c}$ & old & with reacceleration\\
& (-, 5) &  \\
\enddata
\tablenotetext{a}{Intensity in $\mu G$ of the ordered and random component respectively (additional details on B-field parameters are in \ref{sec:model}).}
\tablenotetext{b}{This is the only model that fits synchrotron data, as in O2018.}
\tablenotetext{c}{These are similar to the standard models used for gamma-ray analyses, as in \cite{diffuse2}. DRE$\_$comb is the most similar to those propagation models.}
\label{table1}
\end{deluxetable}

 All the models have the same gas, interstellar radiation field, and CR source distribution. Adopting more complex 3D gas distributions as in \cite{Gulli} do not affect our results, because our comparison is made between models with the same gas distribution. The same consideration applies to different CR source distributions and interstellar radiation field.
In our models the typical energy density of the magnetic field is comparable to the total energy density of the photons (starlight, infrared, and CMB). This brings to similar importance of the IC emission and the synchrotron emission in the energy budget of the Milky Way, as found in \cite{Strong2010}.\\ 
The reference model, PDDE, is the only one fitting synchrotron data. To this model, three models are compared: PDDEBold uses an old B-field that do not fit synchrotron data (spatially and in intensity); DRE does not fit synchrotron spectrum because of the high secondary electron and positron CR spectrum due to reacceleration processes; DRE$\_$comb combines the old B-field (of PDDEBold), with the high CR electron and positron spectrum (of DRE). DRE$\_$comb has very similar condition of the models used in \cite{diffuse2} and still officially and largely adopted for {\it Fermi} LAT studies.

\begin{table}
\begin{center}
\caption{The table shows the propagation parameters of the models. The description of each parameter can be found in the text.}
\begin{tabular}{lccccc}
 \hline
 \hline
\\
         Propagation  & PDDE \&   & DRE \& \\ 
          parameters & PDDEBold  & DRE$\_$comb \\                
  \\
 \hline
\\
  D$_{0}$ $^{(a)}$ (cm$^2$~s$^{-1}$)& 12.3 & 14.6    \\
  D$_{br}$ $(GV)$& 4.8 & -   \\
  $\delta_{1}$ & -0.64 & 0.33\\
  $\delta_{2}$ &0.58 & -   \\
 V$_{Alf}$ (Km~s$^{-1}$)& - & 42.2  \\
 z (kpc) & 4 & 4 \\
 \hline
\label{table2}
\end{tabular}
\end{center}
$^{a}$ D$_{xx}$=10$^{28} \beta D_0(R/ D_R)^\delta$ cm$^2$ s$^{-1}$, with $D_R$=40GV. \\
\end{table}

Table~1 summarizes the general properties of the models. Additional details on the CR parameters can be found in Table~\ref{table2} and in O2018, and references therein. Propagation parameters in Table~\ref{table2} are: $D_{0_{xx}}$, the normalization of the diffusion coefficient at the reference rigidity $D_R$; $D_{br}$, the rigidity break where the index of the rigidity can assume different values ($\delta_1$ and $\delta_2$); the Alfven velocity $V_{Alf}$, and the halo size z.

Below we report more details on the four models used.
\subsubsection{PDDE}\label{sec:model}
This is a pure diffusion model with no reacceleration processes. The propagation parameters are taken as in O2018, where the electron spectrum is fitted to AMS02, Voyager I, and synchrotron data. The intensity of the ordered component of the B-field is fitted to the Planck polarization map \citep{LFPlanck,CompPlanck}, while the intensity of the random component is fitted to the 408 MHz map \citep{haslam1981, Remazeilles}. 
The B-field formulation is from \cite{sun2008} and \cite{sun2010} model as used in \cite{Orlando2013} and \cite{PlanckBfield}, but with intensities refitted to the synchrotron data, due to the different electron spectrum with respect to the works above. 
This B-field model contains a disk, a halo, and a toroidal component for the ordered B-field, and it is less complex and sophisticated than \cite{JF} model. It was also the best model in \cite{Orlando2013}, used as well in \cite{PlanckBfield}, and it was found \cite{PlanckBfield} to reasonably reproduce the large-scale Planck synchrotron and dust maps, as well as the \cite{JF} model.

In more details, the B-field formulation for the random component is a simple exponential law: \\
$B$$_{ran}$ = $B_0$$_{ran}$ $exp(-(R-R$$_{sun}$ $)/R$$_0$$_{ran}$) $exp(-|z|)/z$$_0$$_{ran}$);\\
where $B_0$$_{ran}$=~4.9~$\mu G$, $z_0$$_{ran}$=~4~kpc, and $R_0$$_{ran}$=~30~kpc.
For the ordered disc and halo components 
we use the B-field formulations as in \cite{sun2008} and \cite{sun2010}. 
For the disc field we took their ASS model
plus reversals in rings (ASS+RING), with $B_0$$_{disk}$= 2.7 $\mu G$, while for the halo field we took the toroidal component with $B_0$$_{halo}$=~2.7~$\mu G$. B-field models are in 3D. 
Additional details on the PDDE model can be found in O2018.

It is worthy noting that our updated ordered B-field intensity is similar to the intensity of the regular B-field obtained from rotation measurements  \citep[see e.g.][and references therein]{Beck}. This means that the anisotropic, or striated, component of the B-field is negligible or very low with respect to previous works \citep{JF,Jaffe2013, Orlando2013, Ferriere, PlanckBfield}. This is due to the larger density of electrons measured by AMS02 and Pamela with respect to previous measurements. A dedicated work is in preparation regarding synchrotron studies only, in line with \cite{Orlando2013} and \cite{PlanckBfield}.  

\subsubsection{PDDEBold}
This is a pure diffusion model with no reacceleration processes. Propagation parameters are similar to O2018 and the PDDE model. The electron spectrum is fitted to AMS02 and Voyager I data. The B-field formulation for the random component follows the same
analytic exponential form as in the PDDE, but with different
parameter values such as $B_0$$_{ran}$=~5~$\mu G$, $z_0$$_{ran}$=~2~kpc, and $R_0$$_{ran}$=~10~kpc, as in \cite{diffuse2}. The intensity of the B-field are not reproducing synchrotron data. The B-field formulation is in 2D, instead of 3D as in PDDE model, and no ordered component is accounted for, similarly to \cite{diffuse2}. This model does not fit synchrotron data and it is used in the present work with the PDDE model to present the effect of the B-field only.

\subsubsection{DRE}
This is a standard reacceleration model similar to the ones used in \cite{diffuse2} for gamma-ray analyses. The propagation parameters are taken as in O2018. The electron spectrum is fitted to AMS02 and Voyager I. The intensity of the ordered component of the B-field is taken as the PDDE model. 
Propagation parameters are listed in Table~\ref{table2}. This model does not fit synchrotron spectral data and it is used in the present work with the PDDE model to present the effect of reacceleration only.

\subsubsection{DRE$\_$comb}
This is a standard reacceleration model similar to the ones used in \cite{diffuse2} for gamma-ray analyses. The electron spectrum is fitted to AMS02 and Voyager I. Differently from PDDE model the B-field intensities were not fitted to synchrotron data. The B-field formulation is the same as used in \cite{diffuse2} and in PDDEBold model. This model does not fit synchrotron spectral data and it is used in the present work with the PDDE model to present the effect of reacceleration and of the B-field combined. 

\section{Results on IC model} \label{sec:results}
To study the differences regarding the calculated IC emission between a model with no synchrotron constraints (PDDEBold, DRE, and DRE$\_$comb), and a model with synchrotron constraints (PDDE), we report the results of the spatial effects on the calculated IC maps with different: 1) B-field models (Section \ref{sec41}); 2) propagation models (with and without reacceleration, Section \ref{sec42}); 3) the combination of B-field models and propagation models (Section \ref{sec43}).\\

\begin{figure}
\centering
\includegraphics[width=0.4\textwidth, angle=0] {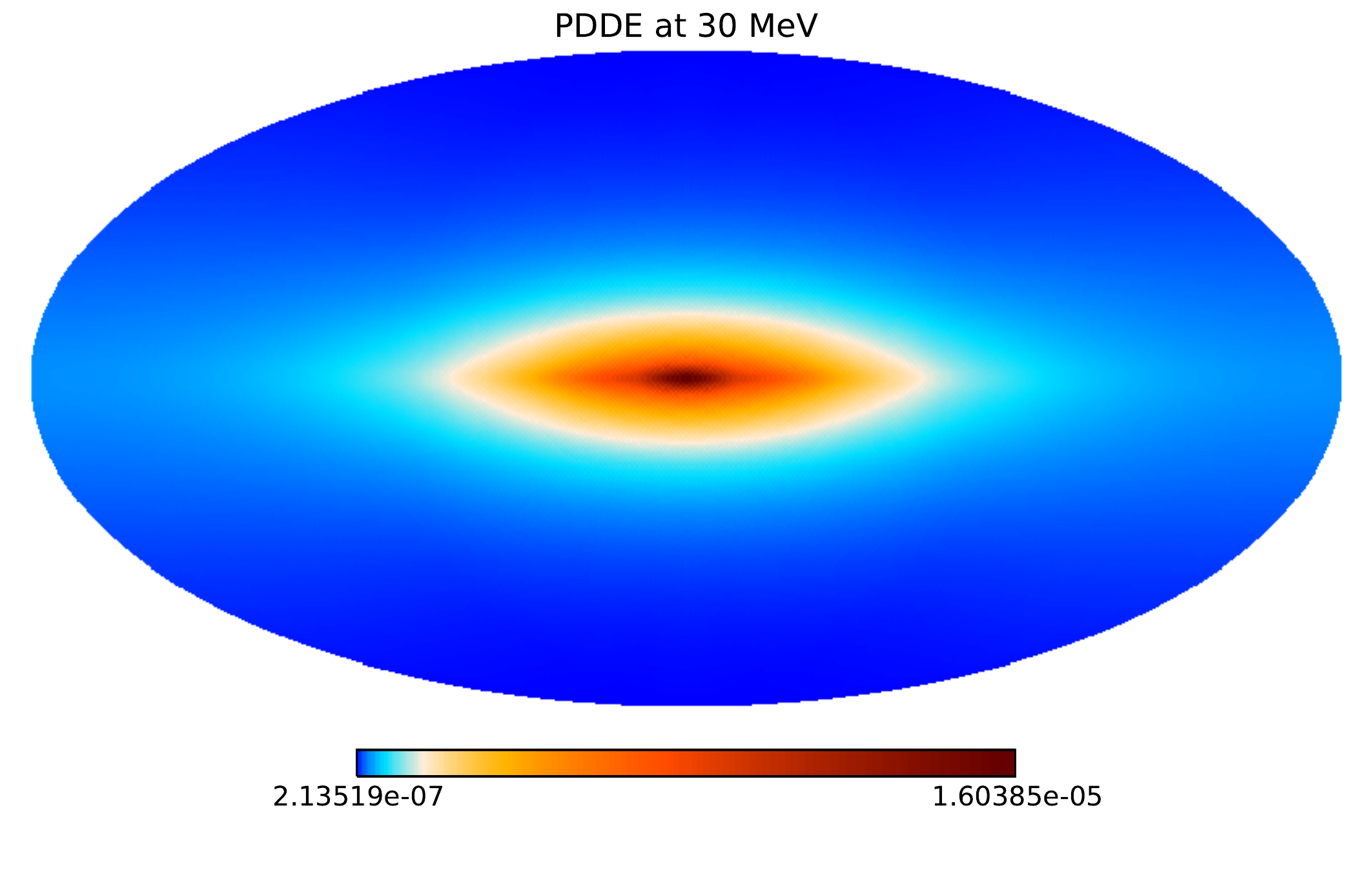}
\includegraphics[width=0.4\textwidth, angle=0] {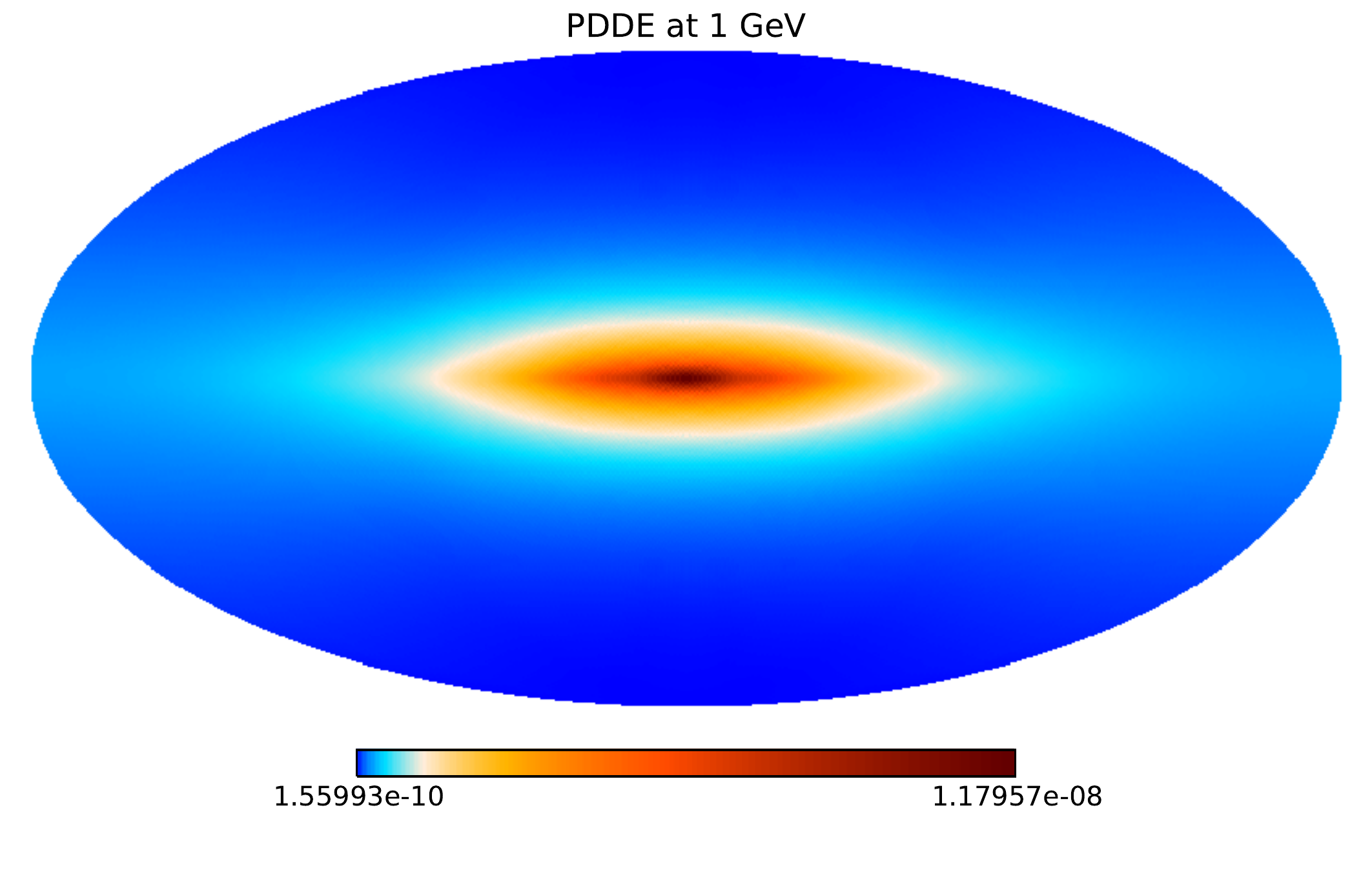}
\includegraphics[width=0.4\textwidth, angle=0] {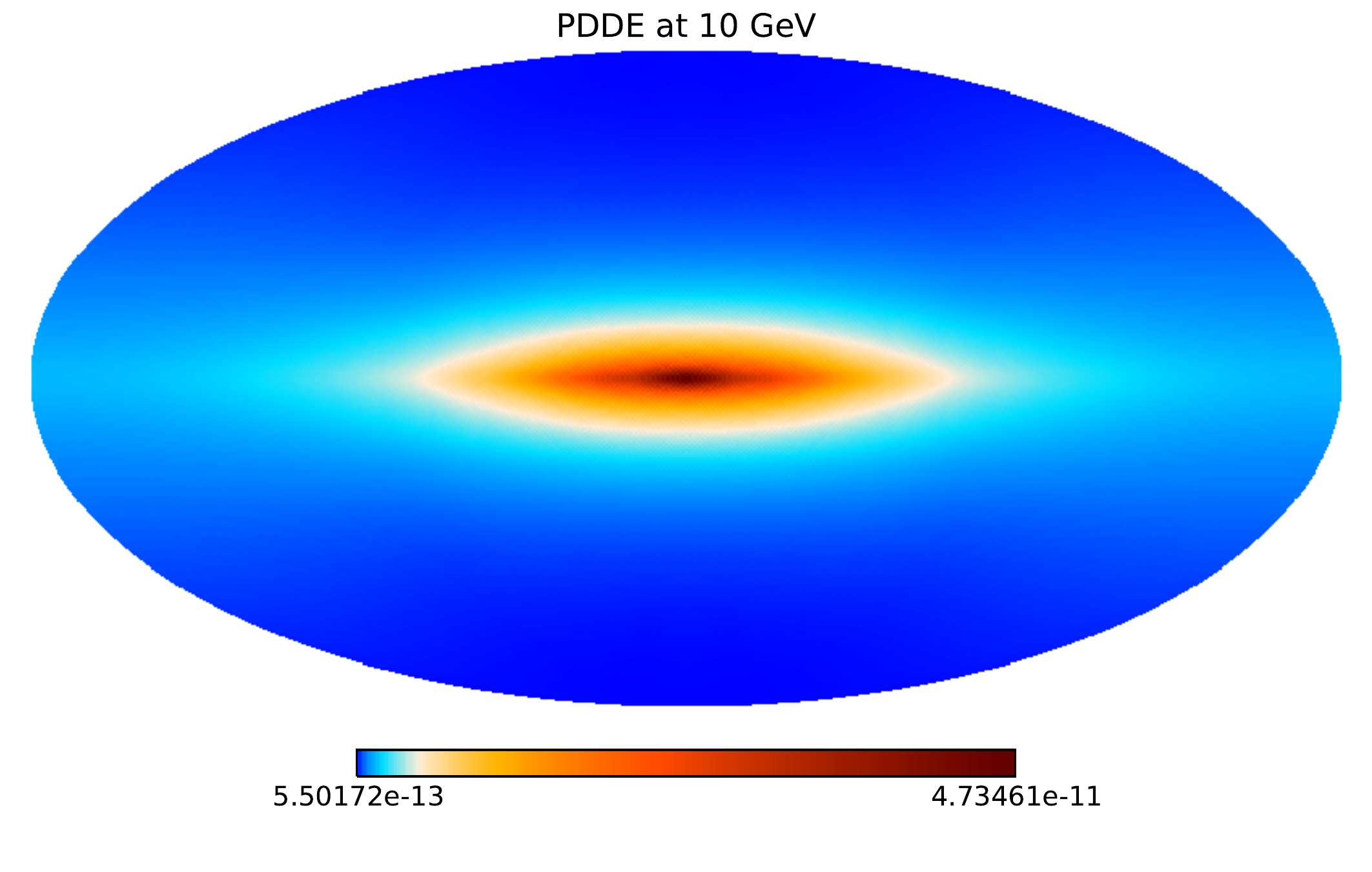}
\caption{Calculated IC emission for PDDE model at 
30 MeV, 1 GeV, 10 GeV top to bottom. Maps are in Galactic coordinates with (l,b)= (0,0) at the center of the map. The colorbar indicates the intensity in units of MeV$^{-1}$cm$^{-2}$sr$^{-1}$s$^{-1}$.}
\label{fig0}
\end{figure}

To verify the effect of including synchrotron constraints
we run GALPROP with the propagation models previously described. All-sky model intensity maps of the IC emission are produced in HEALPix format order 7 \citep{healpix} for different energies.
For illustration Figure \ref{fig0} shows the IC emission component for PDDE model for three energies: 30 MeV, 1 GeV, and 10 GeV.
Then, we calculate the difference between the models for a given energy, taking PDDE as reference model. 
We present the results as all-sky spatial fractional residuals of the calculated IC emission maps between the different models and the reference model. This method of visualization of the differences among models has been extensively applied in the past \cite[e.g.][]{diffuse2}.
Because here we are interested in the spatial distribution only, the intensity maps of the two different models to be compared are normalized to each other in the entire sky to avoid differences in the normalization of the electron spectra between models. This is important especially for reacceleration models below few GeV, where the density of secondary electron and positrons are large. As illustrative examples, we report results for three given energies (30 MeV, 1 GeV, and 10 GeV), which are covered by the {\it Fermi} LAT and where the large-scale IC emission is important.

\subsection{Effects of B-field models}\label{sec41}

\begin{figure}
\centering
\includegraphics[width=0.4\textwidth, angle=0] {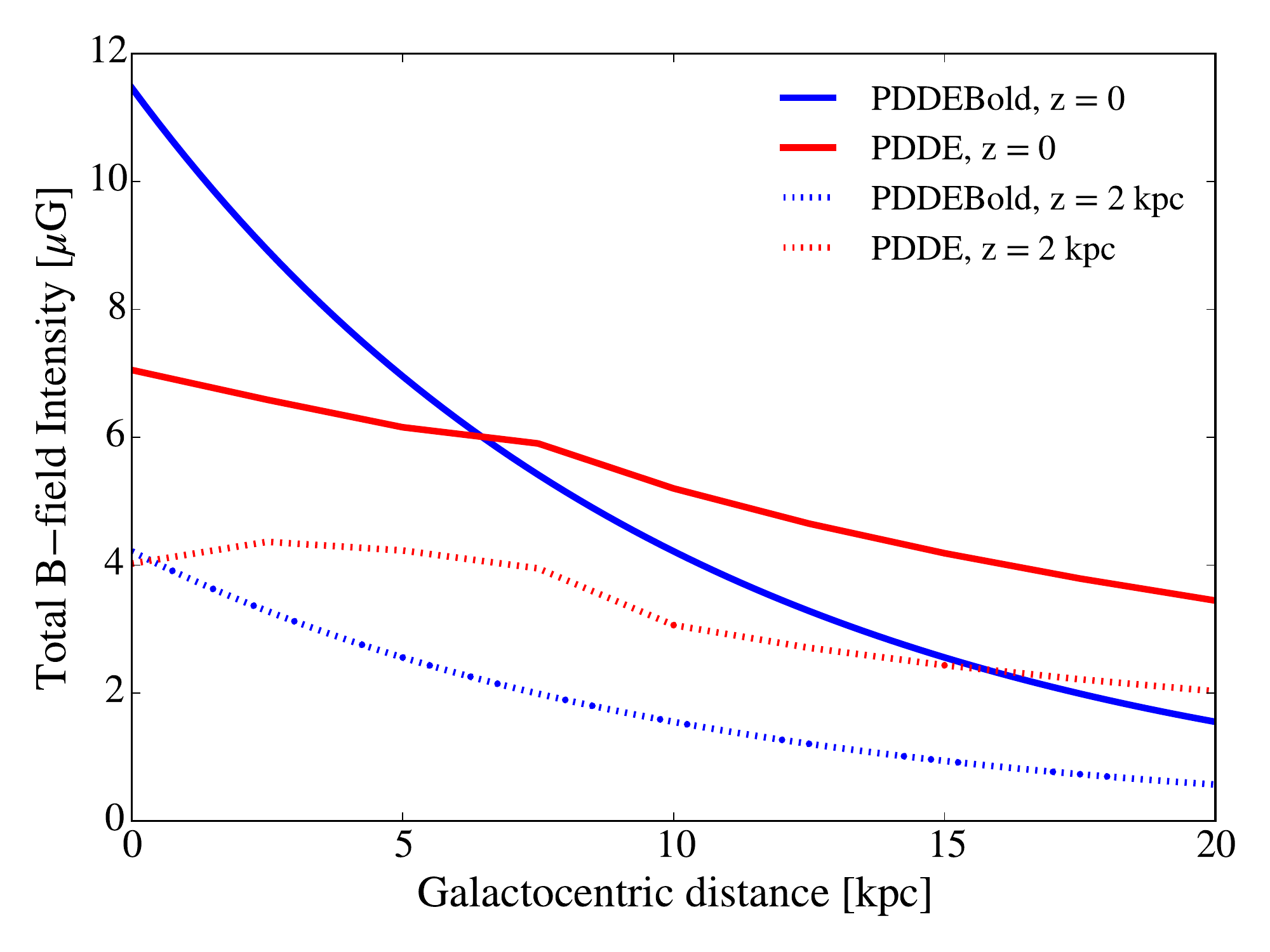}
\includegraphics[width=0.4\textwidth, angle=0] {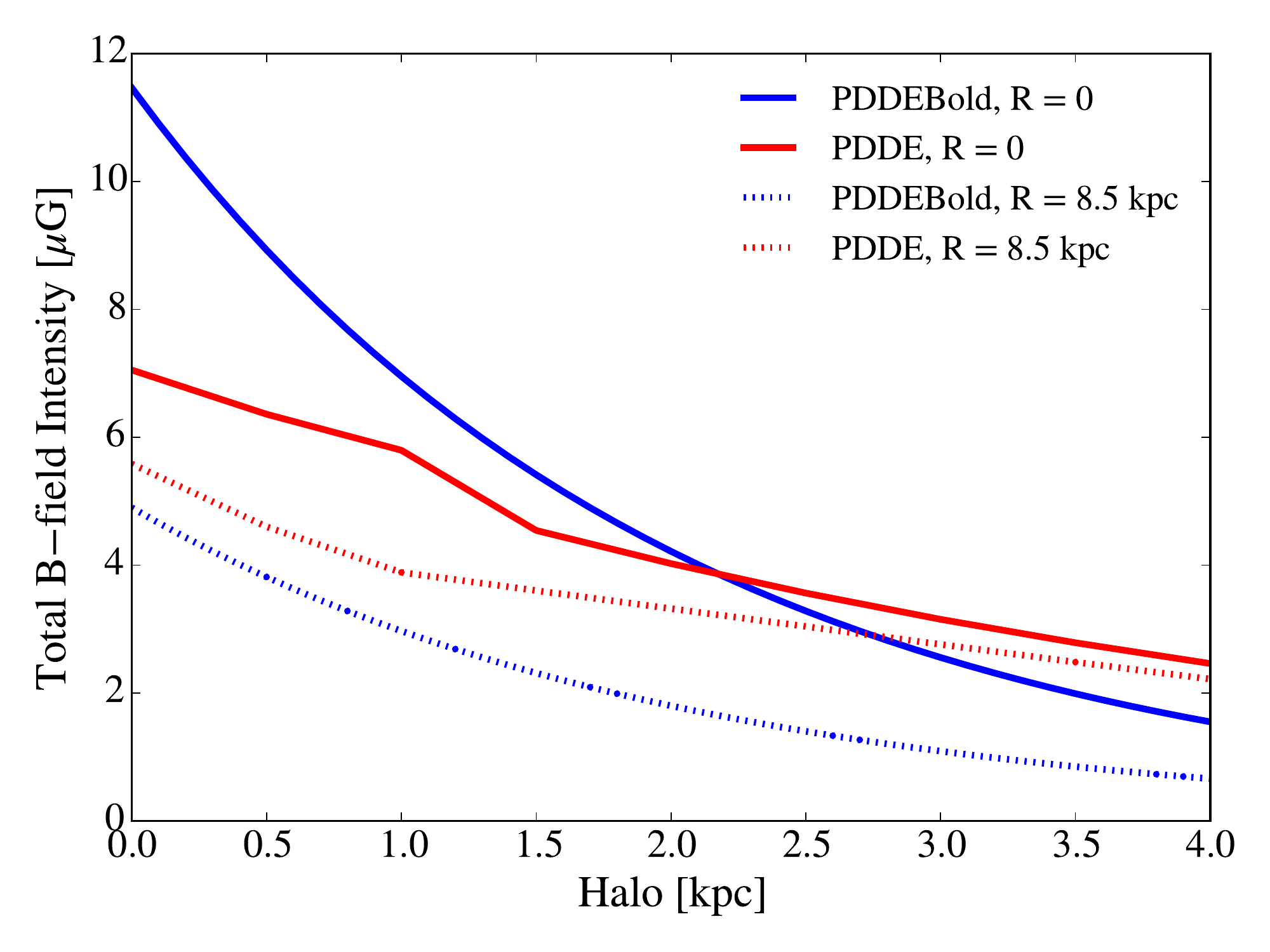}
\caption{{\bf Total B-field intensity for PDDE (red lines) and\ PDDEBold (blue lines) models. {\it Top}: B-field as a function of Galactocentric distance for z = 0 (solid line) and z = 2 kpc (dotted line). {\it Bottom}: B-field as a function of halo hight for R = 0 (solid line) and R = 8.5 kpc (dotted line).}}
\label{fig0b}
\end{figure}

Standard IGE models assume a simple exponential formulation for the B-field, whose intensity and distribution do not fit synchrotron data (spatially or in intensity). PDDEBold model includes a B-field as used in standard models in \cite{diffuse2} and following {\it Fermi} LAT publications, while PDDE model includes ordered and random 3D B-field components that fit synchrotron observations. 
Figure \ref{fig0b} shows the total B-field intensity for PDDE (red lines) and\ PDDEBold (blue lines) models as a function of Galactocentric distance (top) and halo hight (bottom) for a sample of halo hight (z = 0 and z = 2 kpc) and Galactocentric radius (R = 0 and R = 8.5 kpc). While the B-field intensity for model PDDEBold is larger in the inner Galaxy (in the plane), in general for almost the entire Galaxy PDDE has a much larger and flatter B-field intensity. This is not surprising because this model includes also the ordered component (disk and halo), and because the random component has a larger Galactocentric radius and a larger halo hight. More importantly, PDDE has a much uniform B-field intensity over the sky than PDDEBold, which instead is peaked in the inner Galaxy and drops much faster than the PDDE model in the halo. This different gradient between inner Galaxy and halo reflects on the IC spatial maps. 

\begin{figure}
\centering
\includegraphics[width=0.4\textwidth, angle=0] {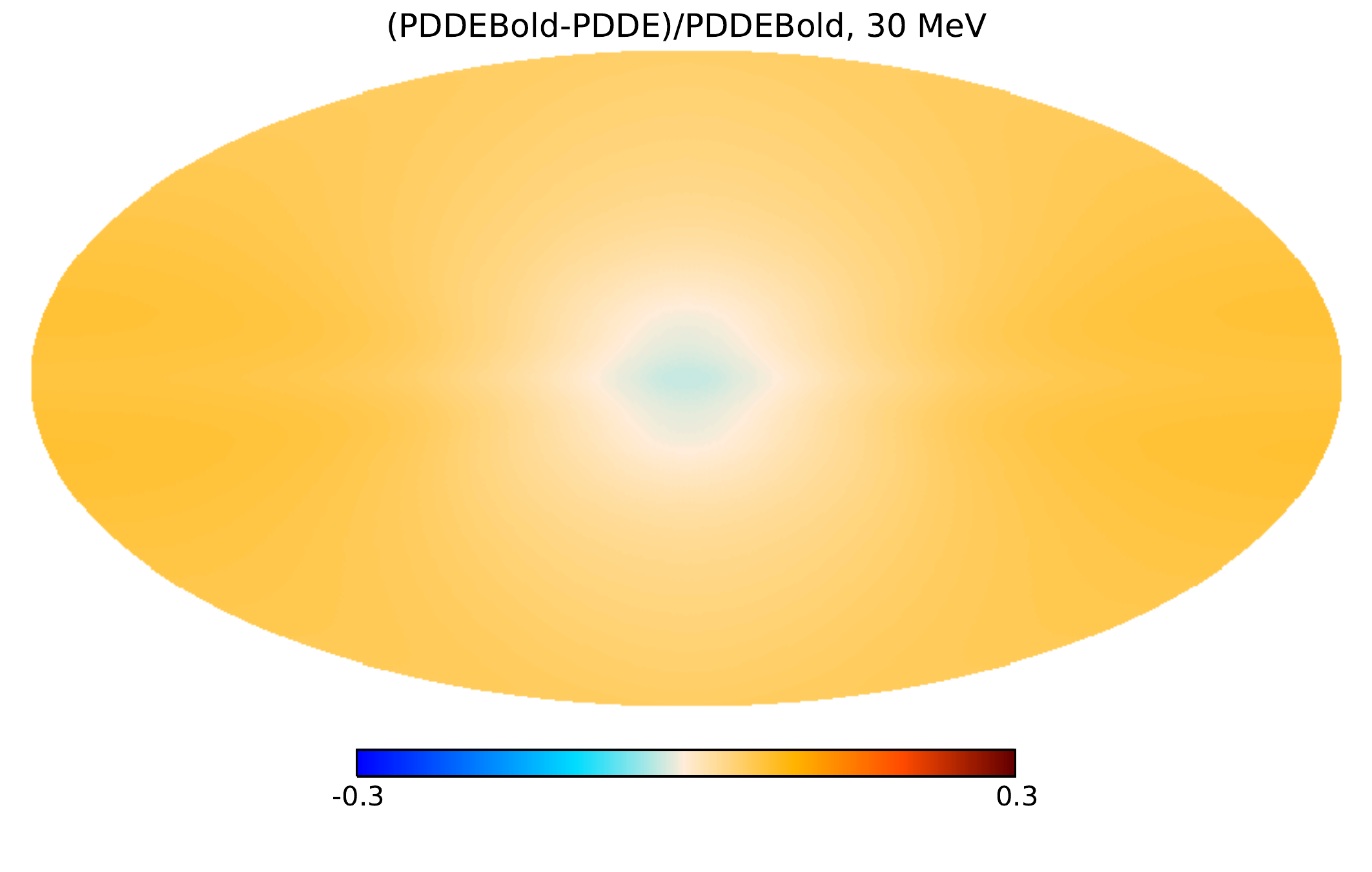}
\includegraphics[width=0.4\textwidth, angle=0] {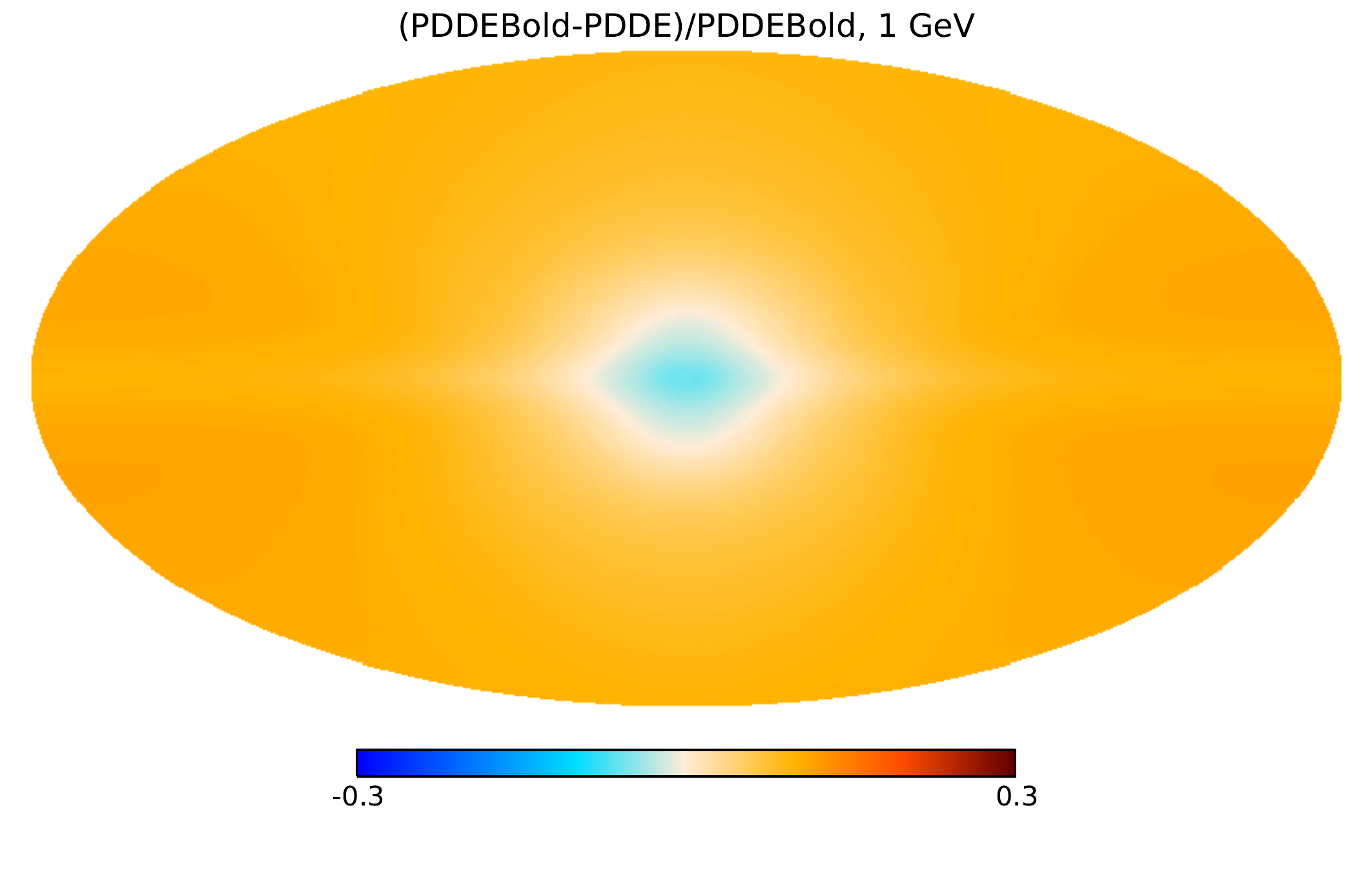}
\includegraphics[width=0.4\textwidth, angle=0] {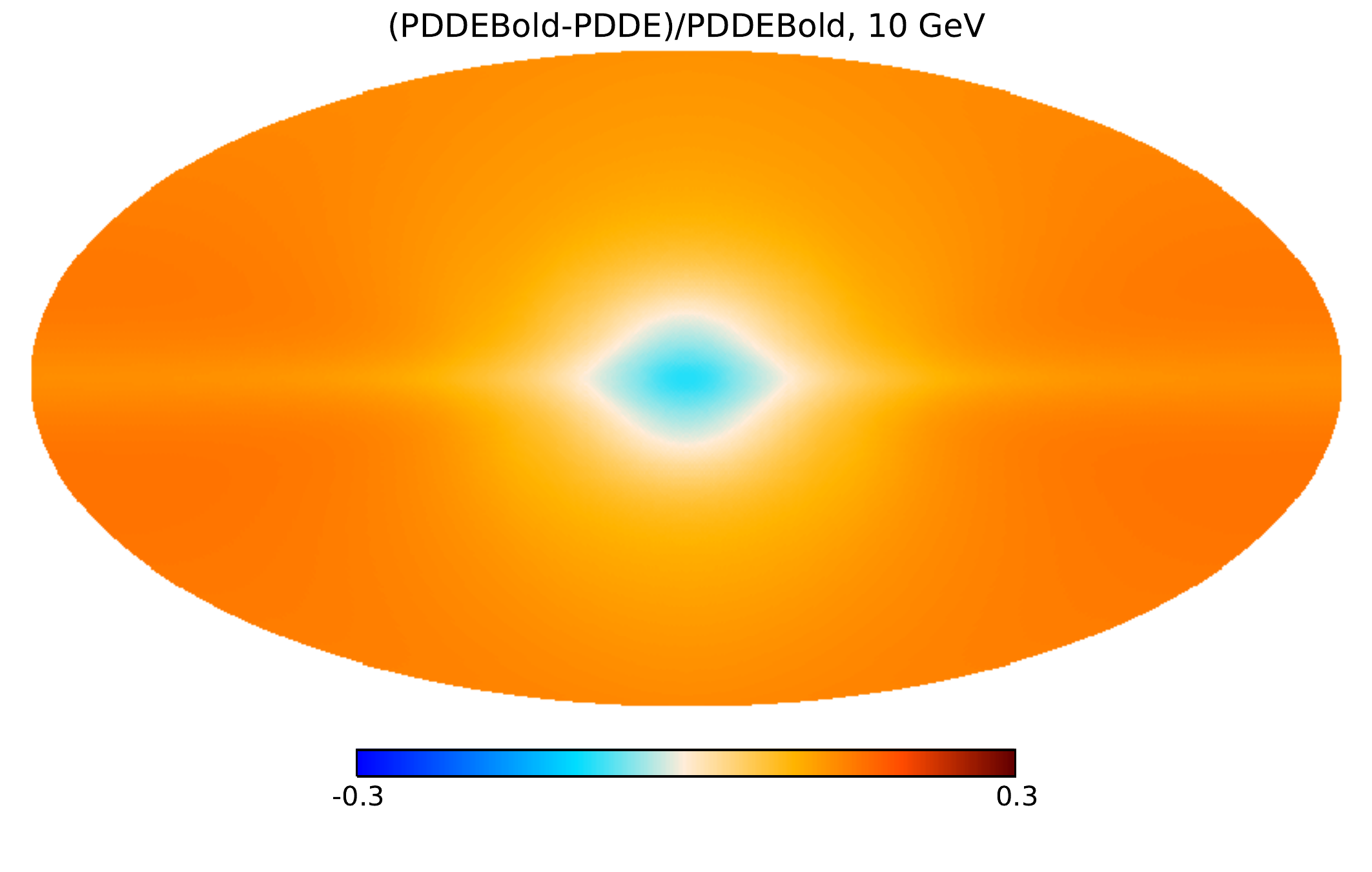}
\caption{All-sky fractional residuals ((PDDEBold-PDDE)/PDDEBold) show the effects of B-field in the calculation of the IC emission. Residuals are reported at 
30 MeV, 1 GeV, 10 GeV, top to bottom. Residual maps are in Galactic coordinates with (l,b)= (0,0) at the center of the map. The colorbar ranges from -0.3 to 0.3.}
\label{fig1}
\end{figure}

Figure \ref{fig1} shows the all-sky spatial fractional residuals of the calculated IC emission between PDDE and PDDEBold model, i.e. (PDDEBold-PDDE)/PDDEBold, for three energies: 30 MeV, 1 GeV, and 10 GeV.  The only difference between PDDEBold and PDDE is the B-field. 
The residuals at all energies exhibit a large-scale trend: the regions outside the inner Galaxy are brighter for the PDDEBold model than for the PDDE model, with respect to the inner Galaxy region. This is due to the different B-field that produces different energy losses, and hence different distribution of electrons in the sky (close and far to the CR sources). Indeed for almost the entire sky the intensity of the total B-field is larger in the PDDE model than in the PDDEBold model. Electrons loose energies faster with a larger B-field, and produce less IC emission in region outside the inner Galaxy. Since energy losses increase with energy, the differences in the IC emission follow the same trend. 
In the inner Galaxy the total B-field intensity is reversed, being larger for the PDDEBold than for PDDE. It is interesting to note that the spatial gradient between regions outside the Galactic plane and the inner Galaxy region increases with the energy, being the IC emission for PDDE more peaked in the inner Galaxy region than for PDDEBold. In other words, the IC intensity from the inner Galaxy to the halo decreases faster for PDDE than for PDDEBold, and this is more significant at higher energies. This exactly reflect the trend of the B-field intensity as shown in Figure \ref{fig0b}.
The extreme intensity values in the calculated IC emission between inner Galaxy and halo reach more than 30\% at 10 GeV.

\subsection{Effects of propagation models}\label{sec42}

\begin{figure}
\centering
\includegraphics[width=0.4\textwidth, angle=0] {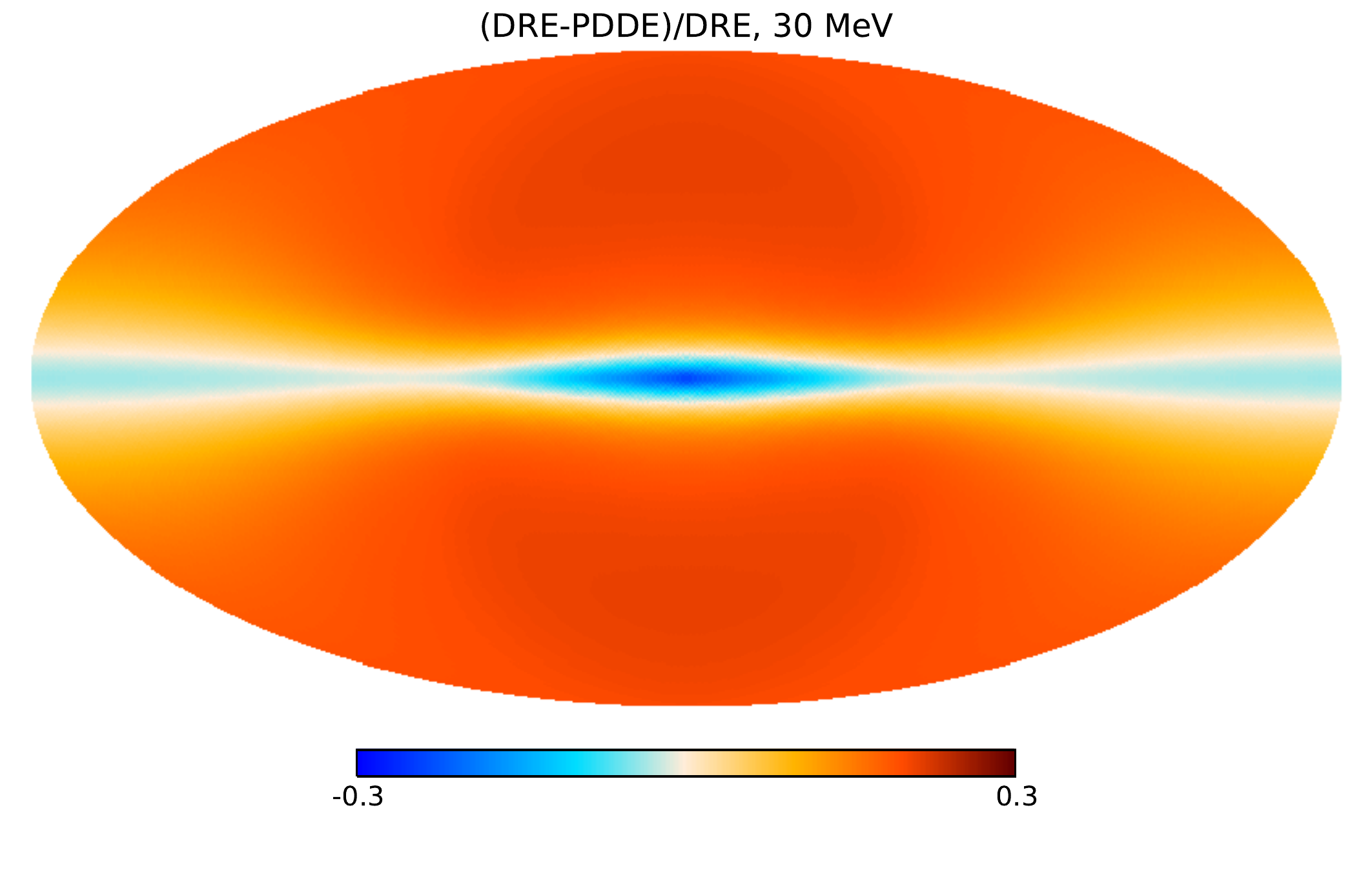}
\includegraphics[width=0.4\textwidth, angle=0] {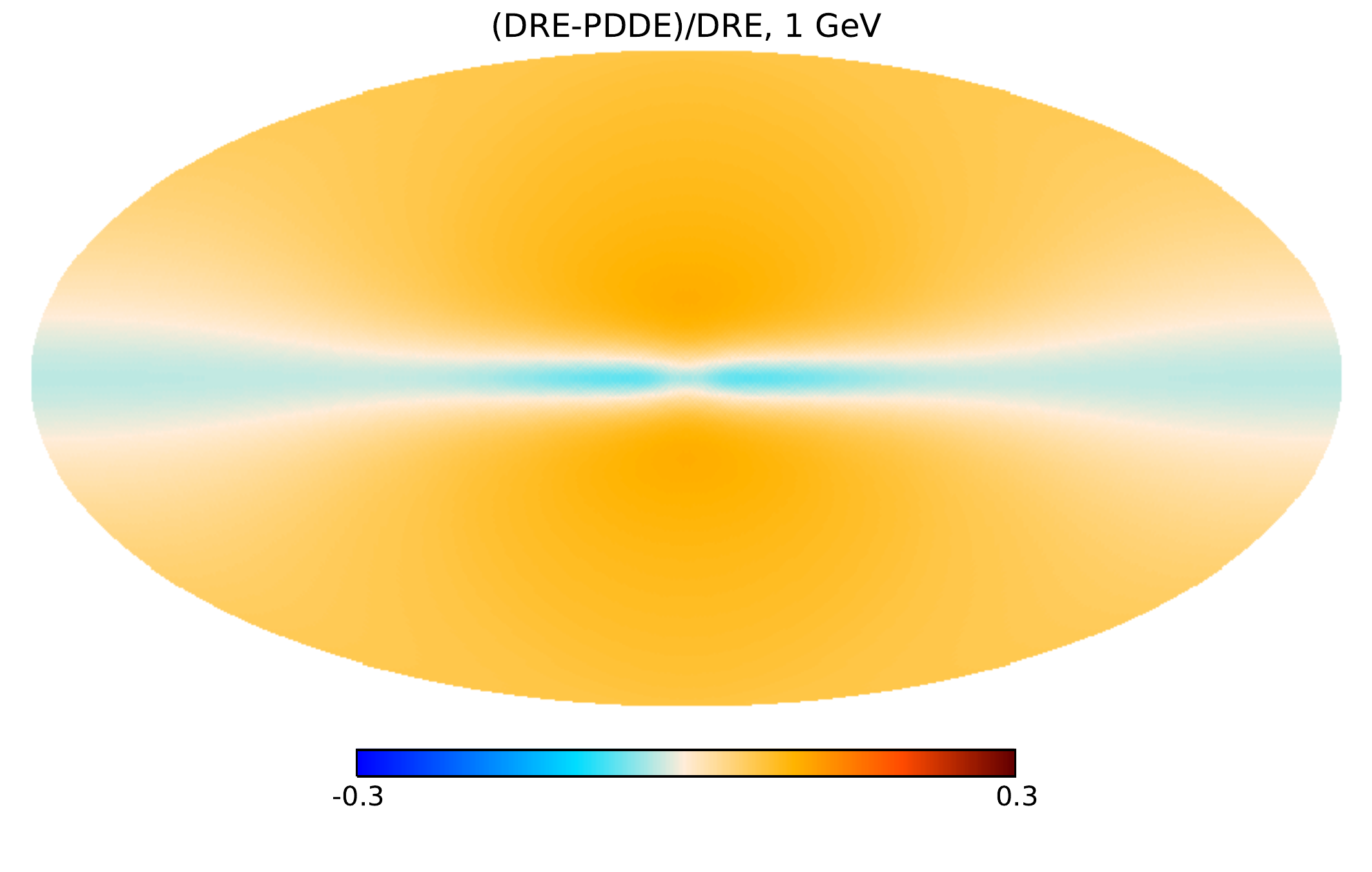}
\includegraphics[width=0.4\textwidth, angle=0] {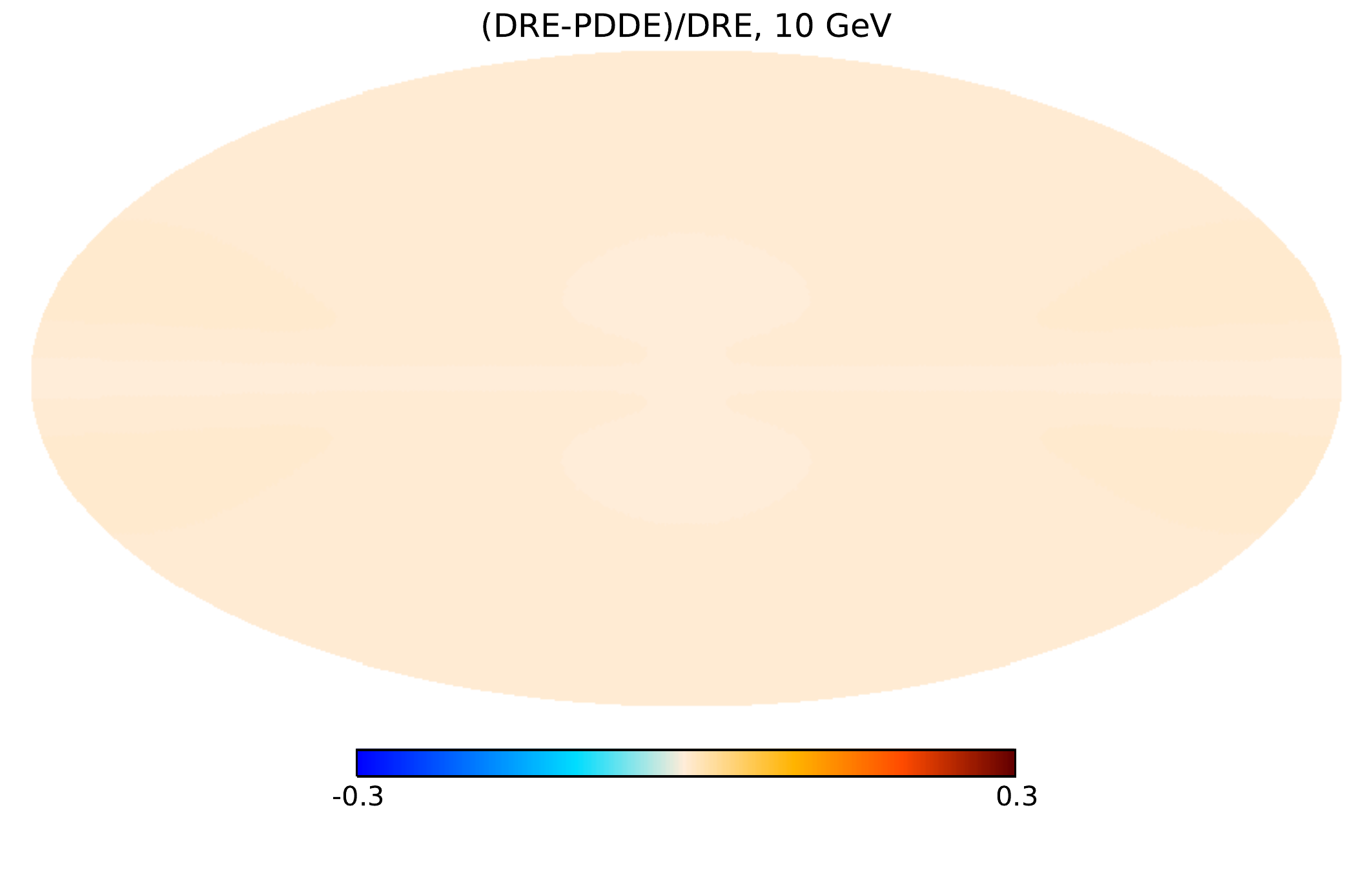}
\caption{All-sky fractional residuals ((DRE-PDDE)/DRE) show the effects of propagation models in the calculation of the IC emission. Residuals are reported at 
30 MeV, 1 GeV, 10 GeV, top to bottom. Residual maps are in Galactic coordinates with (l,b)= (0,0) at the center of the map. The colorbar ranges from -0.3 to 0.3.}
\label{fig2}
\end{figure}

While reaceleration models have been extensively used in gamma-ray and radio data, the all-sky spatial emission of pure diffusion models were used with radio data, and only very recently with gamma-ray data. \\
Figure \ref{fig2} shows the all-sky spatial fractional residuals of the calculated IC emission between PDDE and DRE model, i.e. (DRE-PDDE)/DRE, for three energies: 30 MeV, 1 GeV, and 10 GeV. As already noted above, because here we are interested in the spatial distribution only, the intensity maps of the two different models are normalized to each other in the entire sky to avoid differences in the normalization of the electron spectra between models. This is important especially for reacceleration models below few GeV, where the density of secondary electrons and positrons are large. The spatial difference between PDDE and DRE is given by the different propagation models (reacceleration and pure diffusion models). The B-field is instead the same for the two models. 
DRE is a standard reacceleration model as used in standard models in \cite{diffuse2} and following {\it Fermi} LAT publications, while PDDE does not have reacceleration in order to fit synchrotron spectral observations. 
At the lowest energies (top and middle figures) the intensity from the Galactic plane to the halo decreases faster for the PDDE model than for the DRE model. Indeed, the PDDE model has electrons and positrons much closer to the plane, i.e. where CR source are located, than DRE model. The DRE model, instead, has more electrons and positrons in the halo than PDDE model. This is due to reacceleration processes in the DRE model. Indeed, CRs very
likely spend a considerable time propagating outside the disc in the extended halo.  This produces a similar trend than before: IC emission is much peaked in the inner Galaxy for PDDE than for DRE. It is also interesting to note that the IC emission in the outer Galactocentric radius is significantly brighter for PDDE than for DRE model. 
These differences are more notable at the lowest energies (maps at the top), where reacceleration processes have the most effect. Above 10 GeV (map at the bottom) this effect is negligible. \\
The extreme intensity values in the calculated IC emission between inner Galaxy and halo reach $\sim$60\% at 30 MeV.

\subsection{Effects of B-field and propagation models combined}\label{sec43}

\begin{figure}
\centering
\includegraphics[width=0.4\textwidth, angle=0] {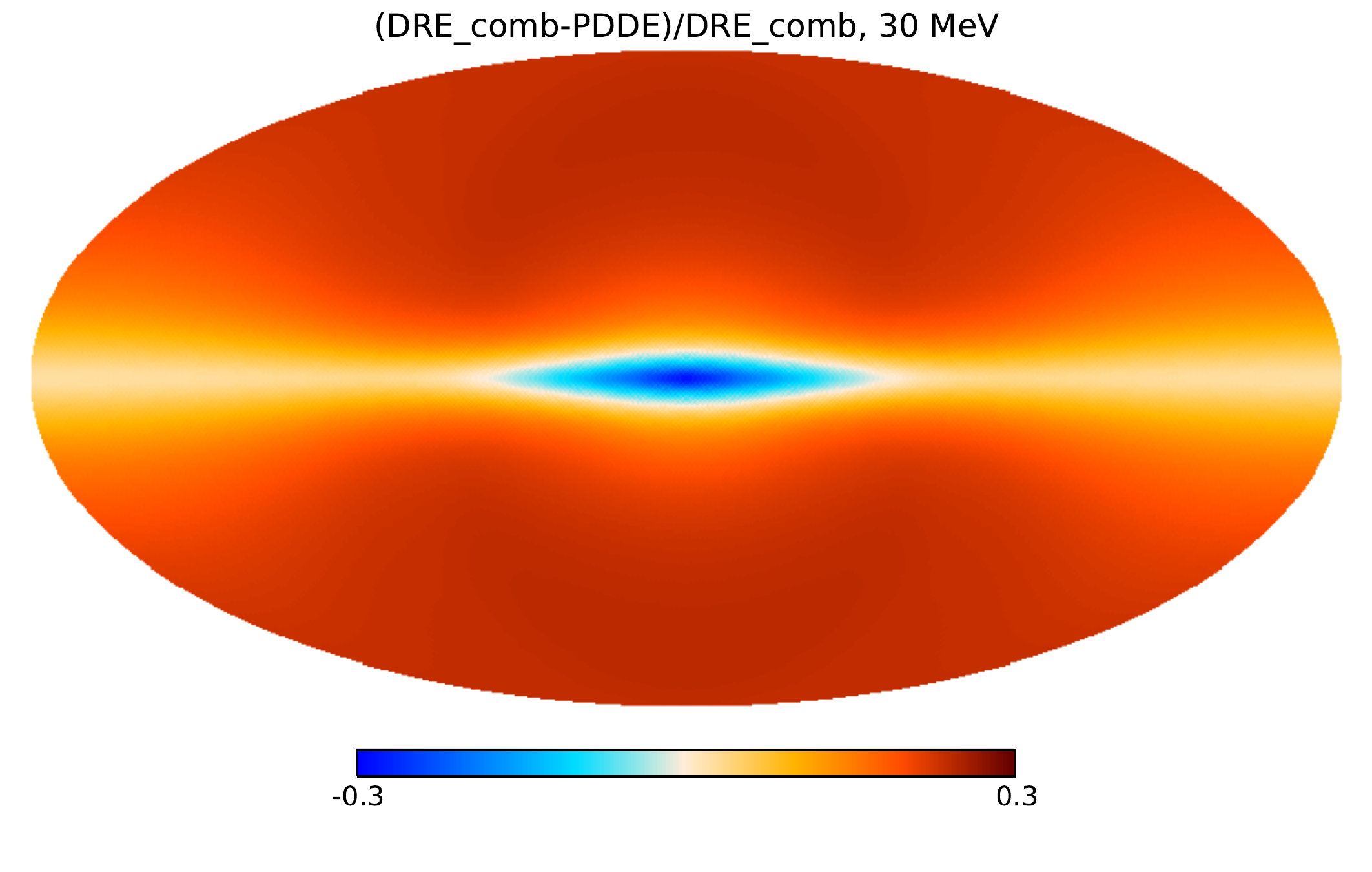}
\includegraphics[width=0.4\textwidth, angle=0] {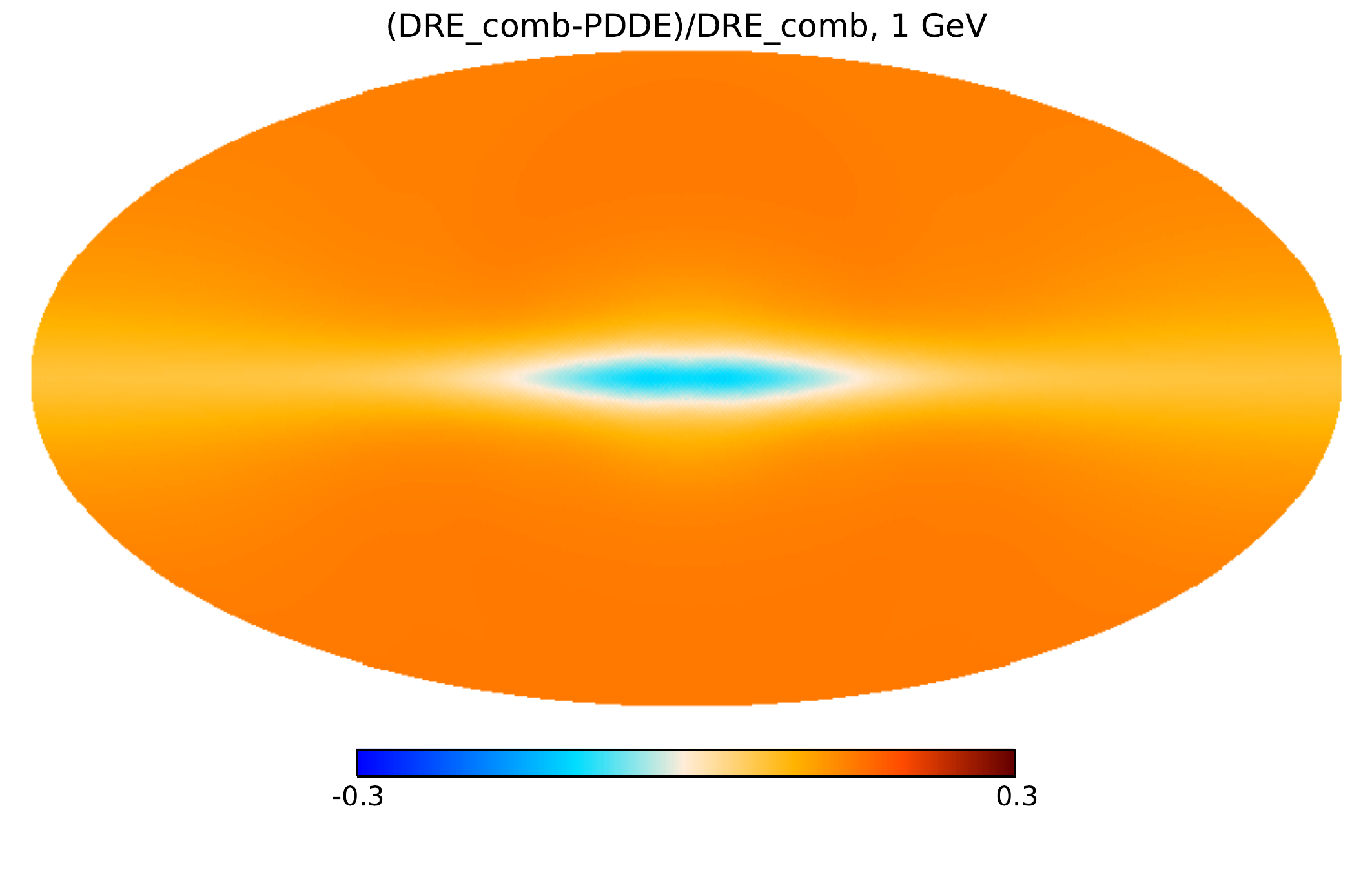}
\includegraphics[width=0.4\textwidth, angle=0] {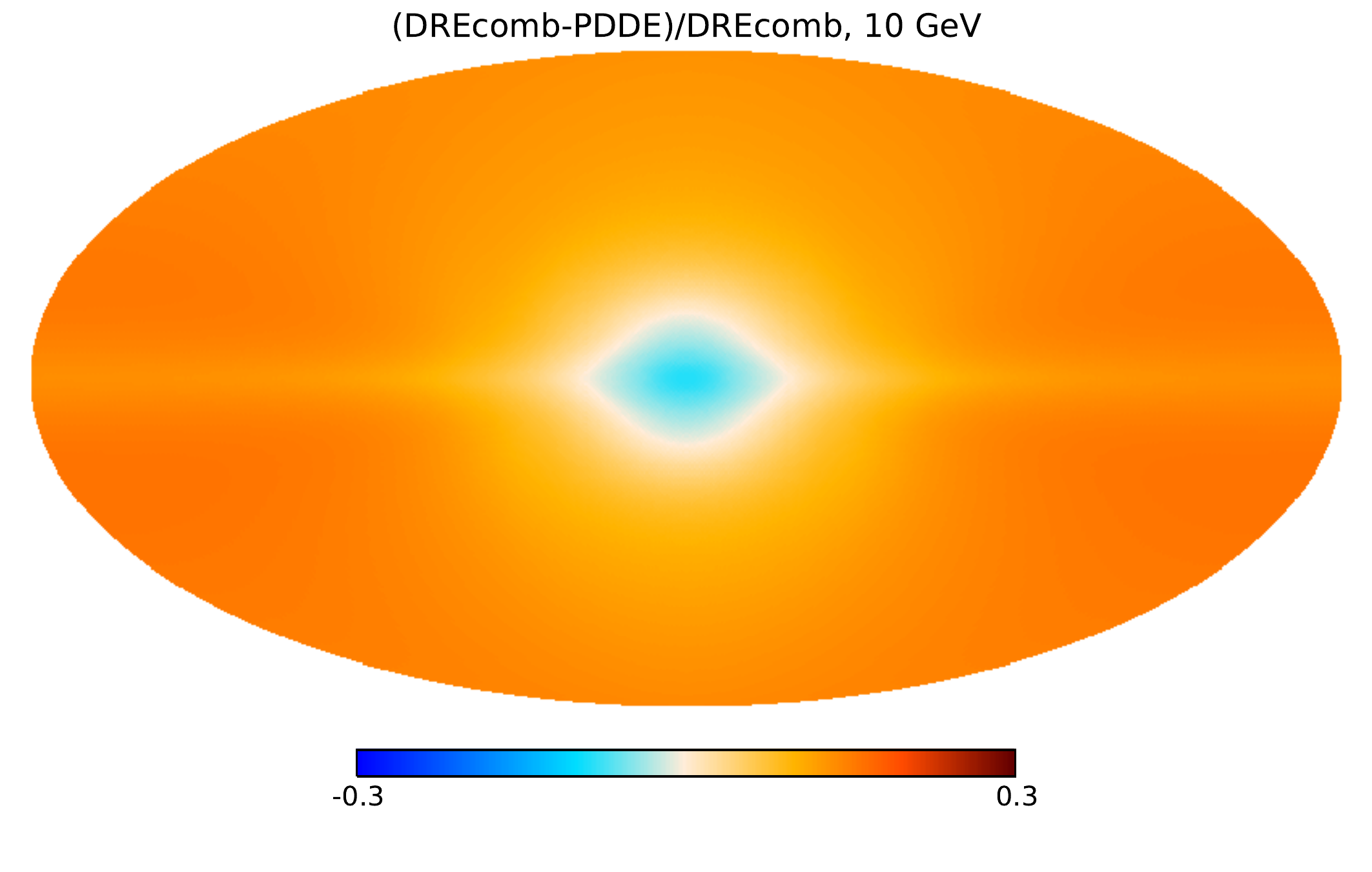}
\caption{All-sky fractional residuals ((DRE$\_$comb-PDDE)/DRE$\_$comb) show the effects of propagation models in the calculation of the IC emission. Residuals are reported at 
30 MeV, 1 GeV, 10 GeV, top to bottom. Residual maps are in Galactic coordinates with (l,b)= (0,0) at the center of the map. The colorbar ranges from -0.3 to 0.3}
\label{fig3}
\end{figure}

Figure \ref{fig3} shows the all-sky spatial fractional residuals of the calculated IC emission between PDDE and DRE$\_$comb model, i.e. (DRE$\_$comb-PDDE)/DRE$\_$comb, for three energies: 30 MeV, 1 GeV, and 10 GeV.  The spatial difference between PDDE and DRE$\_$comb is given by the presence and the absence of reacceleration processes and by the different B-field.  
DRE$\_$comb is a standard reacceleration model as used in standard models in \cite{diffuse2} and following {\it Fermi} LAT publications, which does not account for synchrotron observations.

In general the combination of the updated B-field and propagation model strengthen the differences in the residuals maps.
It is interesting to point out that the PDDE model provides a significantly more peaked IC emission in the inner Galaxy region than DRE$\_$comb model at all energies.
In general the extreme intensity values in the calculated IC emission between inner Galaxy and halo for the two models reach the order of $\sim$60\%.

\section{Discussions and Conclusions}
Usual analyses of {\it Fermi} LAT data are based on IC model maps from propagation models that are fitted to the data as they are, or rarely they are fitted in Galactocentric rings, as in more recent sophisticated analyses.
There is a common consensus about the need of realistic models of the IC emission. Indeed, any advancement in the precision of the IGE models provides a huge impact on our understanding of present gamma-ray observations and can offer insights in describing gamma-ray features seen with {\it Fermi} LAT.
Usual propagation model maps adopted in most of the gamma-ray analyses neglect the information coming from interstellar synchrotron radiation in radio and microwaves without taking advantage of them. In particular from synchrotron studies two outcomes are of fundamental importance for models of the IGE IC emission: the knowledge of the B-field (its intensity and spatial distribution), and the information of propagation models. In more detail, the 2D B-field formulation used in standard models for {\it Fermi} LAT analyses does not contain any halo and large-scale ordered components, which have a 3D formulation. More importantly, the B-field intensity, which is fundamental for calculating propagation and energy losses, is not constrained in those models. In addition, usual IGE models adopted in most of the gamma-ray analyses, are propagation models with reacceleration, which have been found to be strongly disfavored by radio data.
This study presents the effects on the IC gamma-ray spatial model maps if results from synchrotron studies are included in the propagation modeling.
We investigated the effects on the IC component calculated with the GALPROP propagation code with no reacceleration and with updated 3D B-field models, as supported by synchrotron observations. \\
By investigating these updated models with models that do not contain such results, we found that the spatial effect in the all-sky IC maps is of the order of $\pm$30$\%$. Such an effect applies to any CR source distribution, gas, and interstellar radiation field model used. 
Remarkably, the propagation model with a B-field that accounts for synchrotron data produces more peaked IC emission in the inner Galaxy than the standard models used as references for gamma-ray analyses. The difference between the halo and the inner Galaxy reaches even 30\%, increasing with energy. It is interesting to note that a recent analysis of {\it Fermi} LAT data based on standard models \citep{P7IG} found an excess of data over models that was best fitted with an enhanced IC component due to enhancements of either the interstellar radiation field or the CR electrons, given their degeneracy. Some other works attribute instead this excess to a population of unresolved sources, to a dark matter component, or to an anisotropic diffusion coefficient. In this work we found that a correct B-field model could accommodate this enhanced IC in the inner Galaxy. 
In addition, for the same B-field model, propagation models that do not have racceleration produces more IC emission in the plane and especially at large Galactocentric distances than standard models used as reference for gamma-ray analyses. A larger IC emission in the outer plane might be related to the observed excess of data over present IGE models \citep[e.g.][]{outer2010, outer2011, diffuse2, Evoli2012}, which would need more gamma rays than predicted at large Galactocentric distances, and that has been attributed to a higher CR density than predicted, additional gas, or to an anisotropic diffusion coefficient. \\    
In conclusion, updated models, as suggested here, provide a more realistic basis for physical interpretation of the gamma-ray data and some features that are not represented by present models, such as the excess in the inner Galaxy and at large Galactocentric radius. 
We conclude that accurate IGE models can be derived with a multifrequency approach in a self consistent way. 
Model parameters are provided, which work with synchrotron data and CR measurements and that can be used for CR propagation models in studies at gamma-ray energies. \\
This has an important impact for future studies not only with {\it Fermi} LAT, but also with possible forthcoming mission at MeV-GeV, such as AMEGO \citep{amego}, e-ASTROGAM \citep{eastrogamWB} (and All-Sky ASTROGAM), and GAMMA-400 \citep{gamma400}.

\section*{Acknowledgements} 
E. Orlando acknowledges support from NASA Grants No. NNX16AF27G. \\
Useful comments by the anonymous reviewer are acknowledged.\\
This work makes use of HEALPix\footnote{http://healpix.jpl.nasa.gov/} described in \cite{healpix}.

\end{document}